\documentclass[aps,pra,reprint,groupedaddress]{revtex4-1}
\usepackage{graphicx}
\usepackage{dcolumn}
\usepackage{bm}
\newcommand{\dd}{\mathrm{d}}
\newcommand{\topbaseline}[1]{%
\raisebox{\dimexpr\ht\strutbox-\totalheight\relax}{#1}}
\begin{document}


\title{XUV frequency comb metrology on the ground state of helium}


\author{Dominik Z. Kandula}
\altaffiliation{Present Address: Max-Born-Institut, Max-Born-Stra\ss e 2A, 12489 Berlin}

\author{Christoph Gohle}
\altaffiliation{Present Address: Ludwig-Maximilians-Universit\"at M\"unchen, Schellingstrasse 4, 80799 M\"unchen}

\author{Tjeerd J. Pinkert}
\author{Wim Ubachs}
\author{Kjeld S.E. Eikema}
\email{k.s.e.eikema@vu.nl}
\affiliation{LaserLaB Amsterdam, Institute for Lasers, Life and Biophotonics, VU University, De Boelelaan 1081, 1081HV Amsterdam, The Netherlands}



\date{\today}

\begin{abstract}

The operation of a frequency comb at extreme ultraviolet (XUV) wavelengths 
based on pair-wise amplification and nonlinear upconversion 
to the $15^{th}$ harmonic  of pulses from a frequency comb laser in the 
near-infrared range is reported. Following a first account of the experiment [Kandula et al.,~Phys.~Rev.~Lett.~{\bf 105}, 063001 (2010)], an extensive review is given of the demonstration that the resulting 
spectrum at 51 nm is fully phase coherent and can be applied to precision 
metrology. The pulses are used in a scheme of direct-frequency-comb 
excitation of helium atoms from the ground state to the $1s4p$ and 
$1s5p$ $^1$P$_1$ states. Laser ionization by auxiliary $1064$~nm pulses 
is used to detect the excited state population, resulting in a cosine-like 
signal as a function of the repetition rate of the frequency comb with a 
modulation contrast of up to 55\%. Analysis of the visibility of this 
comb structure, thereby using the helium atom as a precision phase ruler, 
yields an estimated timing jitter between the two upconverted comb laser 
pulses of 50 attoseconds, which is equivalent to a phase jitter of $0.38(6)$ 
cycles in the XUV at $51$~nm. This sets a quantitative figure of merit for 
the operation of the XUV comb, and indicates that extension to even 
shorter wavelengths should be feasible.  The helium metrology 
investigation results in transition frequencies of $5740806993(10)$~MHz 
and $5814248672(6)$~MHz for excitation of the $1s4p$ and $1s5p$ $^1$P$_1$ 
states, respectively. This constitutes the first absolute frequency 
measurement in the XUV, attaining unprecedented accuracy in this windowless 
part of the electromagnetic spectrum. From the measured transition 
frequencies an eight-fold improved $^4$He ionization energy of $5945204212(6)$~MHz 
is derived. Also a new value for the $^4$He ground state Lamb shift is 
found of $41247(6)$~MHz. This experimental value is in agreement with recent 
theoretical calculations up to order $m\alpha^6$ and $m^2/M\alpha^5$, but 
with a six times higher precision, therewith providing a stringent test 
of quantum electrodynamics in bound two-electron systems.

\end{abstract}


\maketitle

\section{Introduction}
Atomic spectroscopy has been paramount for the discovery of the laws of 
physics. The ordering of spectral lines in the hydrogen 
atom has led to the Rydberg formula and the concept of quantization 
in the old Bohr model led eventually to the formulation of quantum 
mechanics. In 1947 further advance 
was made with the observation by Lamb and Retherford that the $2 ^{2}\mathrm{S}_{1/2}$ 
and $2 ^{2}\mathrm{P}_{1/2}$ states in hydrogen are not degenerate, but 
differ by 1 GHz~\cite{Lamb1947p241,Lamb1950p549}. This result 
lead to the development of quantum electrodynamics 
(QED)~\cite{Feynman1949p769,Schwinger1948p1439,Tomonaga1946p27}, 
which is the most precisely tested physics theory to date. 
Since the first measurement by Lamb and Retherford, QED 
contributions to energy levels in atoms are referred to as ``Lamb
shifts''. The theory has been elaborately tested and confirmed by experiments
in various bound systems, ranging from atomic hydrogen~\cite{Huber1999p1844,Parthey2010p233001},
via one-electron heavy-ion systems~\cite{Gumberidze2005p223001},
exotic atomic systems such as positronium~\cite{Hagena1993p2887} 
and muonium~\cite{Meyer2000p1136} to one-electron molecular ions~\cite{Korobov2009p012501,Koelemeij2007p173002},
while recently a test of the QED has also been reported in a neutral molecule~\cite{Salumbides2011p43005}.
Tests have also been performed on the 
anomalous magnetic dipole moment of the electron (``g-2"), so 
that in fact QED can now be used to derive 
a new value for the fine structure constant from such a measurement~\cite{Hanneke2008p120801}. 

The two main QED contributions to the energy of an atomic system are
virtual photon interactions ("self energy") and screening of the nuclear 
charge due to the creation (and annihilation) of electron-positron pairs ("vacuum polarization"). The
magnitude of these effects depends on the considered system and its energy
eigenstate, and is studied by means of precision spectroscopy. In atoms 
such as hydrogen and helium, the strongest QED effects are observed in 
the electronic ground states. For this reason spectroscopy involving the ground 
state is preferred. In the case of hydrogen this is achieved by inducing 
a two-photon transition at 2 $\times$ 243 nm from the $\mathrm{1S}$ 
ground state to the $\mathrm{2S}$ excited state, which has reached 
a level of precision that allows a detailed comparison between theory 
and experiment~\cite{Fischer2004p230802}. This 
comparison is currently limited by the uncertainty in the measured proton 
charge radius~\cite{Pachucki2003p113005,Fischer2004p230802,Pohl2010p213},
which in itself is not a QED phenomenon. 
Assuming that the QED calculations are correct, one can  
also use the $\mathrm{1S-2S}$ experiment in hydrogen to determine 
the effective proton charge radius. In this respect it is interesting 
to note that a recent experiment with muonic hydrogen (where the 
electron in hydrogen is replaced by a muon), has resulted in a 
proton size that differs 5$\sigma$~\cite{Pohl2010p213} with the 
CODATA value that is based mainly on hydrogen spectroscopy. The 
origin of this difference is still under debate~\cite{Jentschura2011p7,DeRujula2011p26,Cloet2011p012201,Barger2011p153001}.

In principle a more stringent test of QED could be performed by 
experiments using atoms with a higher nuclear charge $Z$, as the 
non-trivial QED effects scale with $Z^4$ and higher. For this 
reason experiments have been performed on many different high-$Z$
ionic species such as $\mathrm{U}^{91+}$~\cite{Gumberidze2005p223001}. 
However, the required (very) short wavelengths for excitation, 
such as hard-X-rays, makes it very difficult to perform absolute 
frequency measurements (see e.g.~\cite{Epp2010p194008} and references therein).

Even for $Z=2$ in the He$^+$ ion or the neutral helium atom, where QED effects 
are at least 16 times larger than in hydrogen, it remains 
difficult to perform high resolution spectroscopy. In the 
case of He$^+$ a two-photon excitation from the ground 
state requires 60 nm, while for neutral helium 120 nm is 
required to drive the $1s^{2}$~$^1\mathrm{S}_0 - 1s2s$~$^1\mathrm{S}_0$ 
two-photon transition. 
One-photon transitions require even shorter wavelengths, 
in the extreme ultraviolet (XUV), e.g. $58$~nm to excite 
the $1s^2$~$^1\mathrm{S}_0 - 1s2p$~$^{1}\mathrm{P}_{1}$ 
first resonance line in neutral helium.
 
A major obstacle to precision spectroscopy in the 
XUV is the lack of continuous wave (CW) narrow bandwidth 
laser radiation. Instead pulsed-laser techniques have been 
used. The first laser-based measurement of the $1s^2$~$^1\mathrm{S}_0 - 1s2p$~$^{1}\mathrm{P}_{1}$ 
resonance line at $58.4$~nm was achieved via upconversion 
of the output of a grating-based pulsed dye laser~\cite{Eikema1993p1690}.
Subsequently, pulsed amplification of CW lasers and 
harmonic generation in crystals and gases led to production 
of narrower bandwidth XUV radiation. In this fashion spectroscopy on neutral 
helium has been performed from the ground state using 
ns-timescale \emph{single} pulses of 58.4~nm~\cite{Eikema1997p1866,Eikema1996p1216}. 
In an alternative scheme two-photon laser excitation of neutral 
helium at 120 nm was achieved~\cite{Bergeson1998p3475,Bergeson2000p1599}. 
In both experiments transient effects resulted in 
"frequency chirping" of the generated XUV pulses, 
which limited the accuracy of the spectroscopy to about 50 MHz. 

Here we overcome this problem by exciting transitions 
using a \emph{pair} of phase coherent XUV pulses, 
produced by amplification and high-harmonic generation 
(HHG) of pulses from a frequency comb laser. In such a scheme
most of the nonlinear phase shifts and those due to short-lived 
transients cancel as only differential pulse 
distortions enter the spectroscopic signal. In effect, 
a frequency comb in the XUV is generated. The employed 
method of spectroscopy with this XUV comb laser is a 
form of Ramsey spectroscopy~\cite{Ramsey1949p996}. 
Early experiments using phase-coherent 
pulse excitation in the visible part of the spectrum 
used a delay line~\cite{Salour1977p757}, a resonator~\cite{Teets1977p760} 
or modelocked lasers~\cite{Eckstein1978p847}
to create two or more phase-coherent pulses. More recently, 
amplified ultrafast pulses have been split using a 
Michelson interferometer or other optical means to 
create two coherent XUV pulses after HHG~\cite{Salieres1999p5483,Bellini2001p1010,Kovacev2005p223903}. 
Coherent excitation of argon in the XUV with a delay 
of up to $\approx$100 ps~\cite{Liontos2010p832} has 
been demonstrated in this manner. However, with these 
methods no absolute calibration in the XUV has been 
demonstrated up to now as it is very difficult to 
calibrate the time delay and phase difference between 
the pulses with sufficient accuracy. In the present experiment we 
use a frequency comb laser (FCL)~\cite{Diddams2000p5102,Holzwarth2000p2264} to obtain phase-coherent 
pulses in the XUV which allows for a much higher 
resolution and immediate absolute calibration.


Single femtosecond laser pulses can be made sufficiently intense 
for convenient upconversion to XUV or even soft
x-ray frequencies using HHG~\cite{Seres2005p596}. 
However, as a consequence of the Fourier principle, 
the spectral bandwidth of such pulses is so large that it inhibits
high spectral resolution. Frequency combs combine high resolution 
with high peak power by generating a continuous \emph{train} of 
femtosecond pulses. In this case the spectrum of the pulse train 
exhibits narrow spectral components (modes) within the spectral 
envelope determined by the spectrum of a single pulse. The modes 
are equally spaced in frequency at positions given by: 
\begin{equation}\label{comb_frequencies}
f_{n} = nf_{rep} + f_{CEO}
\end{equation}
were $n$ denotes the integer mode
number, $f_{rep}$ is the repetition frequency of the pulses, and $f_{CEO}$
is the carrier-envelope offset frequency. The latter relates to the
pulse-to-pulse phase shift $\Delta\phi_{CE}$ between the carrier wave
and the pulse envelope (carrier-envelope phase or CEP) by 
\begin{equation}\label{phase-shift}
\Delta\phi_{CE} = 2\pi f_{CEO}/f_{rep}.
\end{equation}
Each of the comb modes can be used as a high-resolution probe
almost as if it originated from a continuous single-frequency 
laser~\cite{Fendel2007p701,Marian2005p023001,Wolf2009p223901}. If the entire pulse 
train can be phase coherently upconverted, the generated
harmonics of the central laser frequency should exhibit a similar spectrum
with comb frequencies $mf_{rep} + qf_{CEO}$ , where $m$ denotes again
an integer mode number, and $q$ is the integer harmonic order under
consideration. By amplification of a few pulses from the train, and
producing low harmonics in crystals and gases, an upconverted
comb structure has been demonstrated down to $125$~nm~\cite{Witte2005p400,Zinkstok2006p061801}. However, to reach
wavelengths below $120$~nm, HHG has to be employed requiring
nonlinear interaction at higher intensities in the non-perturbative
regime~\cite{Lewenstein1994p2117}. It is well established that HHG can be phase coherent to
some degree~\cite{Salieres1999p5483,Bellini2001p1010,Cavalieri2002p133002,Lewenstein1994p2117,Zerne1997p1006,Liontos2010p832}, and
attempts have been made to generate frequency combs based
on upconversion of all pulses at full repetition rate~\cite{Gohle2005p234,Jones2005p193201,Ozawa2008p253901,Yost2009p815}. 
However, due to the low XUV-pulse energies the comb
structure could so far not be verified in the 
XUV-domain for those sources.
That limitation can be overcome by combining parametric amplification 
of two frequency comb pulses in combination with harmonic upconversion. 
This was recently demonstrated with direct frequency comb excitation at 
$51$~nm in helium~\cite{Kandula2010p063001}, and in this paper a 
full and detailed description is given of that experiment. 

The article is organized as follows: 
in section II the measurement principle is explained, followed in section III 
with a description of the different parts of the experimental setup. The 
general measurement procedure and results are presented in section IV. In 
section V, part A, a discussion of all systematic effects that 
need to be taken into account to determine the ground state ionization 
potential from the measured transitions frequencies is given. This is followed in 
part B with a discussion of the timing jitter in the XUV, which can be 
derived from the measured Ramsey signal. In the 
final section VI the conclusions and an outlook are presented.

\section{Overview and principle of XUV comb generation and excitation}

A frequency comb is normally based on a modelocked laser producing an infinite 
train of pulses with fixed repetition rate and CEP-slip between 
consecutive pulses. The corresponding 
spectrum of such a pulse train consists of a comb of narrow optical modes
associated with frequencies given by Eq.~\ref{comb_frequencies}. 
To convert the frequency comb to the XUV, we select only two pulses from 
the FCL. In this case the spectrum changes to a cosine-modulated continuum, 
but with the peaks of the modulation exactly at the positions of the original FCL spectrum.
This ``broad frequency comb" is converted in a phase-coherent manner to the 
XUV by amplification of the pulse pair to the millijoule level, and subsequent HHG. Once the FCL pulses 
(separated by the time $T = 1/f_{rep}$) are upconverted, they can be used to directly probe 
transitions in atoms or molecules. This form of excitation with two
pulses resembles an optical (XUV) version of the Ramsey method of
spatially (and temporally) separated oscillatory 
fields~\cite{Ramsey1949p996,Witte2005p400}. In the present case, the interacting fields are 
not separated in space, but only in time. Excitation with two (nearly) identical pulses
produces a signal which is cosine-modulated according to:
\begin{equation}\label{broad_comb_signal}
S(T)\propto cos(2\pi(f_{tr}T ) - \Delta\phi(f_{tr}))
\end{equation}
when varying $T$ 
through adjustment of $f_{rep}$ of the comb laser. In this 
expression $f_{tr}$ is the
transition frequency and $\Delta\phi(f_{tr})$ is the spectral phase
difference at the transition frequency between the two pulses. 
Relation~(\ref{broad_comb_signal}) is valid only for weak interaction, 
which is the case in the current experiment given an excitation probability of $\ll 1$ per atom.    

Without HHG and in the absence of additional pulse distortions the phase shift
$\Delta\phi := \Delta\phi(f_{tr})$ is equal to $\Delta\phi_{CE}$. The
excitation signal will exhibit maxima corresponding to those frequencies where the
modes of the original comb laser come into resonance with the
transition. 

The spectral phase difference of the generated XUV pulse 
pair cannot be determined directly. Therefore we need to 
propagate the spectral phase difference from the frequency 
comb through the parametric amplifier and the HHG process 
into the interaction region. The phase shift $\Delta \psi(f)$ 
imprinted on the pulses by the non-collinear optical parametric double-pulse
amplifier (NOPCPA) is measured direcly using 
an interferometric technique described previously~\cite{Kandula2008p7071}. 
To model the  HHG process we employ a slowly varying envelope 
approximation described in section~\ref{discussion}, which 
yields a differential  XUV phase shift of the form
\begin{equation}\label{phase_shift_q}
\Delta\phi_{q} = q(\Delta\phi_{CE} + \overline{\Delta\psi}) + \Delta\psi_{q}, 
\end{equation}
where $\overline{\Delta \psi}$ denotes the carrier envelope phase 
accumulated in the NOPCPA (compare Sec.~\ref{measurement}), $q$ is the harmonic order of the resonant radiation 
and $\Delta\psi_q$ is an additional phase shift due to nonlinear and 
transient response in the HHG process~\cite{Lewenstein1995p4747,Pirri2007p138} and transient 
effects such as ionization. Note that only differences in
phase distortion between the two subsequent pulses from the FCL affect
$\Delta\phi_{q}$ and therefore the Ramsey signal. Shared distortions, 
such as frequency chirping due to uncompensated time-independent 
dispersion, have no influence on the outcome of this experiment.

The frequency accuracy of the method scales with the period of the
modulation ($f_{rep}$, here equal to $100-185$~MHz) rather than the
spectral width of the individual pulses (about $7$~THz in the XUV at
the $15^{th}$ harmonic). 
An error $\delta$ in the value of $\Delta\phi_q$ leads to a 
frequency error in the spectroscopy result of $\Delta f = \delta/(2\pi T)$.
Of the two components contributing to this error, $\Delta\psi(f)$ can 
be expected to be independent of $T$, while the transients contained in
$\Delta\psi_q$ decay with increasing $T$. Therefore $\Delta f$ 
decreases at least with $1/T$, leading to a higher accuracy for 
a longer time separation between the pulses. In practice there is 
still a minimum requirement on the stability and measurement accuracy 
of phase shift of rms 1/200th of a cycle (at the fundamental 
frequency of the FCL in the near infrared). This ensures that the contribution 
$\overline{\delta \psi}$ to the frequency uncertainty of the 
measured transition frequency is small enough so that the 'mode number' 
ambiguity in the transition frequency, due to the periodicity of the signal, can be resolved with confidence.

\begin{figure}[!h]
 \includegraphics[width=0.9\columnwidth]{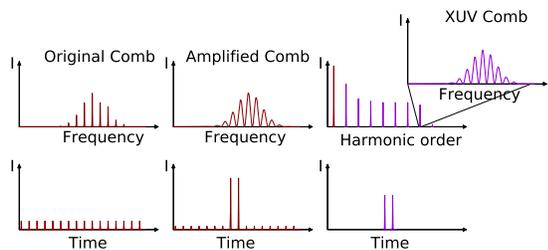}
\caption{Principle of frequency comb generation and spectroscopy in the XUV. (a) Schematic
of the spectral and temporal structure of the generated light at the
different stages in the experiment (left to right): sharp equidistant
frequencies from the infrared FC laser, followed by double-pulse
amplification resulting in a cosine modulated spectrum, and finally
HHG resulting in odd harmonics of the central frequency, where each of
the harmonics consist of a XUV comb in the form of a cosine-modulated
spectrum with period $f_{rep}$.\label{setup}} 
\end{figure}

A sketch of the XUV comb principle in the frequency domain is shown in 
Fig.~\ref{setup}. A FCL serves as a source of phase-coherent pulses. A
pair of such pulses is amplified in a NOPCPA to the mJ level, yielding a cosine modulated spectrum 
when viewed in the frequency domain. The amplified pulses 
are filtered spatially with two pinholes, one placed between
the second and third amplification stage, and one after compression respectively. 
As a result the pulse pairs show less spatial intensity 
and phase variation compared to the unfiltered beam.

Before the IR beam is focused in a krypton jet for HHG, the center of it is 
blocked by a $1.9$~mm diameter copper disk, while the outside is clipped with 
an iris.  The donut-mode shape is used to facilitate the separation 
of the driving IR field from the generated XUV emitted on axis (see also section~\ref{HHG}).

Phase shifts in the NOPCPA, which are not common for both
pulses, and therefore change the position of the comb modes, are measured by means
of spectral interference. A Mach-Zehnder-like configuration is used to interfere 
the original comb pulses with those that are amplified by the NOPCPA, 
as described in~\cite{Kandula2008p7071} and section~\ref{phase}.

\begin{figure}
\includegraphics[width=0.9\columnwidth]{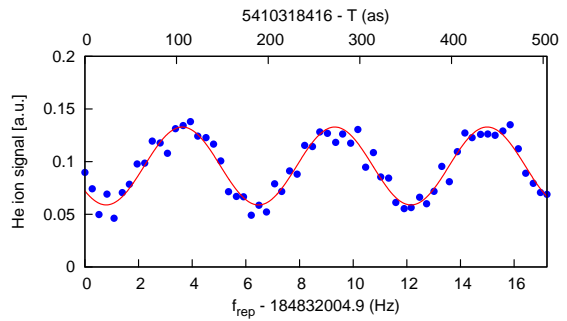}
\caption{The excitation probability of helium at $51.5$~nm on the 
$1s^{2}$~$^{1}\mathrm{S}_{0} - 1s5p$~$^{1}\mathrm{P}_{1}$ transition,
as a function of the repetition
of the frequency comb laser (lower x-axis), and the delay between the
pulses (upper x-axis in attoseconds). In this example $f_{CEO}$ is
locked $46.21$~MHz, and a 1:5 He:Ne mixture is used for the atomic
beam.}
\label{wiggle060515}
\end{figure}

The amplified pulses are focused a few mm
in front of a krypton jet, in which two phase-locked pulses 
of high-order harmonics are generated to create a cosine-modulated comb spectrum in the XUV. 
In order to avoid direct ionization of helium with
higher-order harmonics, the IR intensity in the 
interaction region and the harmonic medium are chosen
such that the desired $15^{th}$ harmonic appears at the cutoff of the
HHG process. The XUV beam crosses a low-divergence
beam of helium atoms perpendicularly, and excites them from the ground state to the
$1snp$~$^{1}\mathrm{P}_1$ state ($n\in\{4,5\}$). A pulse of $1064$~nm radiation
ionizes the excited atoms, which are subsequently detected in a time-of-flight mass
spectrometer. Tuning of the XUV comb is accomplished by changing $f_{rep}$ of the FCL. 
The changing mode separation effectively causes the modes in the XUV to scan over the transition.
An example for a scan of the repetition time corresponding to approximately
$500$~attoseconds (as) is shown in Fig.~\ref{wiggle060515}. 
The number $m$ of the mode that excites the 
measured transition in this experiment is in the order of $50$~million. This means that 
the repetition rate of the fundamental FCL needs to be 
changed only by a few Hz in order to bring an adjacent 
XUV-mode $f_{m\pm1}$ into resonance with a helium 
transition ($f_{m}$ is assumed to be at the transition initially).
As a consequence the ionization signal will be cosine modulated.
In order to resolve the resulting ambiguity in the mode-number assignment, the 
measurement is repeated with different repetition rates, 
corresponding to pulse delays between $5.4$~ns and $10$~ns. 

Besides the frequency-domain perspective, this experiment can 
be viewed also in the time domain. In this case
it can be seen as a pump-probe experiment, which tracks the dynamics
of the electronic wave function that results from mixing the ground state of
helium with the excited $p$-level. The first XUV pulse brings the atom
into a superposition of the ground and the excited state. This
superposition results in a dipole, which oscillates at the transition
frequency with an amplitude that decays according to the lifetime of
the excited state. The second pulse probes this oscillation. 
Two extreme cases can be distinguished. If the 
second pulse is in phase with the dipole oscillation, the
amplitude of the excited state increases, and thus its
detection probability. If the phase of the second pulse is shifted by
$\pi$ with respect to the dipole oscillation of the
helium atom, its oscillatory movement is suppressed and the probability
to find an atom in the excited state is decreased. 

\section{Experimental setup}

\begin{figure}
\includegraphics[width=0.9\columnwidth]{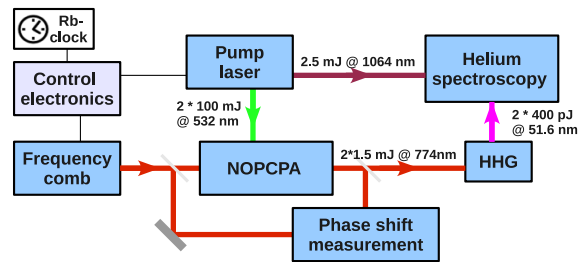}
\caption{Scheme of the experimental setup, 
including: femtosecond frequency comb laser with control electronics, 
non-collinear optical parametric chirped-pulse amplifier (NOPCPA), 
a phase measurement in a Mach-Zehnder interferometer, high harmonic generation (HHG) 
and a spectroscopy apparatus. Part of the pump laser output at
$1064$~nm is used as ionization beam.\label{scheme}}
\end{figure}

The measurement setup consists of four major elements: a frequency comb laser, a
non-collinear parametric amplifier, a phase-measurement system, and a 
vacuum apparatus for HHG and excitation of helium in an atomic beam. 
A schematic overview of the setup is given in Fig.~\ref{scheme}.

\subsection{Frequency-comb oscillator and pulse stretcher}
Phase-coherent pulses are obtained from a home-built Kerr-lens mode-locked Ti:Sapphire frequency
comb laser. The FCL has an adjustable repetition rate between $100$~MHz and
$185$~MHz. Dispersion compensation in the cavity is obtained with a set 
of chirped mirrors, supporting a spectral bandwidth of $60$~nm centered
around $780$~nm. The FCL is stabilized and calibrated against the signal 
of a rubidium clock (Stanford research PRS10), which itself is 
referenced to a GPS receiver so that an accuracy on the
order of $10^{-11}$ is reached after a few seconds of averaging. 
Before sending the FCL pulses to the amplifier, the 
wavelength and bandwidth of the comb pulses is adjusted 
with a movable slit in the Fourier plane of a grating-based
$4-f$~stretcher ($1200$~l/mm, $f=10$~cm).
The bandwidth is set to $6$~nm in this device, ensuring that after upconversion 
to the XUV only one state in helium is excited at a time. The spectral clipping 
and losses from the gratings in the stretcher reduce the pulse energy from about $5$~nJ 
to $60$~pJ. At the same time the added dispersion 
and reduced bandwidth lengthens the pulse to about $2$~ps.

\subsection{Non-collinear optical parametric amplifier}

\begin{figure}[h]
\includegraphics[width=0.9\columnwidth]{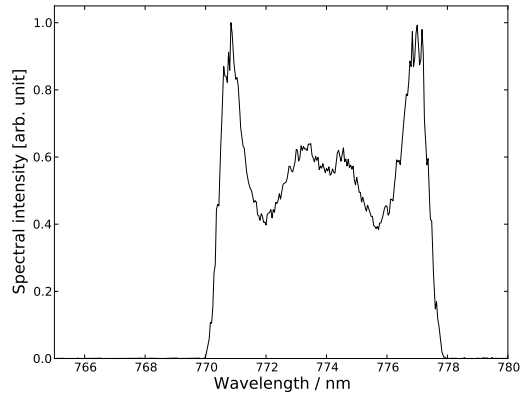}
\caption{A typical spectrum of the pulse-pair after NOPCPA, acquired with 
a resolution of $0.1~\mathrm{nm}$ measured by an Ando AQ-6315A optical spectrum analyzer.
\label{spectrum}}
\end{figure}

A pair of subsequent pulses obtained from the stretcher is amplified in a 
non-collinear optical parametric chirped-pulse amplifier based on two $5$~mm long 
BBO (beta-barium borate) crystals. Here we present a concise description of the system, 
while further details can be found in~\cite{Kandula2008p7071,Witte2006p8168}. The amplifier operates at a 
repetition rate of $28$~Hz, and amplifies two subsequent FC pulses to a level 
of typically $5$~mJ each. The bandwidth of the pulses remains essentially 
unchanged compared to that selected by the slit of the preceding stretcher, although 
saturation effects in the NOPCPA result in a 'cathedral'-like spectrum (see Fig.~\ref{spectrum}). 
The pump light for the NOPCPA (two pulses at $532$~nm, $50$~ps, $80-100$~mJ per pulse) 
is obtained by frequency doubling $1064$~nm light from a Nd:YAG-based pump laser. 
A relay-imaged delay line in the pump laser is used to produce these pulse pairs 
with a time separation between $5.5$~ns and $10$~ns. This time separation is adjusted carefully for 
each value of $f_{rep}$ to match the time delay between consecutive pulses from the FC 
laser at a few-ps level.

Electronic synchronization is employed
(timing jitter less than $1$~ps) so that the pump and comb laser pulses arrive at
the BBO crystals at the same time. The amplification happens in three stages, 
the first two located in the first BBO crystal and the last 
(power amplifier) stage uses the second. Between the
first and second crystal spatial filtering is used to reduce
phase-front errors. After amplification in the NOPCPA, and subsequent compression, a final 
spatial filter is employed (based on a 1:1 telescope with 
$f = 50$~cm lenses and an $80$~micron pinhole in between, mounted in vacuum). 
This reduces wave-front errors and results in a Gaussian 
beam with a diameter of $6$~mm. The pulses of $300$~fs duration are essentially 
diffraction and Fourier-limited with an energy of $1-2$~mJ per pulse.

By choosing parametric amplification, a high gain ($10^{8}$) is achieved while avoiding
some of the transient effects that conventional laser amplifiers
suffer from, such as thermal lensing and inversion depletion. The
reason for this is that in NOPCPA negligible power is dissipated in
the amplifier medium (a nonlinear crystal), and that the amplified
(’signal’) wave is not distorted provided the process is perfectly
phase matched (see e.g. \cite{Renault2007p2363}). Imperfections of 
the wave front of the pump-beam pulses can lead to a spatially-dependent 
phase mismatch in the NOPCPA, which in turn influences the phase of the 
amplified signal beam. Therefore  special attention is paid to match 
the properties (wave front, pulse length, energy, diameter) of the two 
pump pulses so that induced phase effects in the two amplified signal 
pulses are as equal as possible. A Shack-Hartmann sensor was used 
to align their direction within several $\mu$rad and their 
propagation axis within $10~\mu$m.
The remaining spatial differential signal 
distortions are reduced by the spatial filters and finally measured 
interferometrically as described in the next section. 

\subsection{Phase-shift measurements in the IR}\label{phase}

Knowledge of the carrier-envelope phase change between
consecutive pulses of a repetition-frequency stabilized mode-locked
laser is a prerequisite for frequency-comb metrology. Because parametric
amplification influences the phase of the amplified
light~\cite{Ross2002p2945,Witte2007p677}, a method is needed to detect a
differential phase shift between the pulses. This shift is recorded for
each laser shot, using spectral interferometry with the original comb
pulses as a reference in a Mach-Zehnder configuration
(Fig.~\ref{scheme}).

The phase-measurement setup is an advanced version
of the one published earlier~\cite{Kandula2008p7071}. In order to deal with
the demand for higher spectral resolution of the phase-measurement setup and the spatial
dependence of the phase shift between the pulses, several
improvements were made. The most important is a motorized
iris (diameter: $2$~mm), which enables spatial mapping of the differential
phase within 30 seconds, by scanning it along the donut mode (see also section~\ref{HHG}). Such a wave-front
scan is made before every recording of the helium signal as it 
cannot be done for every individual laser shot while 
measuring the helium signal. However, during a
recording of the helium signal the iris is opened and an average value for the shift of $\Delta\phi_{q}$
is monitored for each laser shot. This average is compared with
the wave-front scan made just before a recording of the helium signal 
from which a correction for the average phase shift is
calculated. In this way the effective phase difference change can be
continuously monitored. The differential phase distortion along the donut-mode profile 
has a typical magnitude of $100$~mrad, and a spatial variation of $20-30$~mrad.

Perfect spatial overlap between the amplified and reference pulses 
is ensured by sending both beams through a $2$~meter long large-mode volume (field
diameter $20$~$\mu$m) single-mode photonic fiber. Before doing so, the
pulses are stretched approximately 40 times in a grating stretcher.
The combination ensures that both amplified and reference beam are
perfectly overlapped, while the stretching and the large mode volume 
of the fiber enables to increase the pulse energy for good signal-to-noise 
ratio without inducing self-phase modulation (SPM) in the fiber. This last 
aspect is particularly important because otherwise intensity 
differences between the pulses could induce SPM and corrupt 
the result of the phase shift determination.

\subsection{High harmonic generation}\label{HHG}

The central part of the amplified and spatially filtered beam is
blocked by a $1.9$~mm diameter copper disk to facilitate
separation of the fundamental IR light and generated XUV further downstream. High
harmonics are produced by focusing this donut-shaped beam ($f=500$~mm) 
a few mm in front of a pulsed krypton gas jet. The
intensity at the focus is estimated to be less than $5 \times
10^{13}$~W/cm$^{2}$, whle the local gas density is estimated at a 
few mbar. After the focus the beam encounters an iris of
$0.8$~mm diameter at $40$~cm distance from the jet, which is 
placed in the image plane of the copper disk. The
iris blocks the infrared radiation with an extinction of better than
$1:27$, while the XUV light emitted on axis can freely
propagate. Thereafter the XUV beam passes the interaction chamber
(described below), and enters a normal-incidence focusing grating
monochromator equipped with an electron multiplier to analyze the
harmonic spectrum. We estimate that about $1\times 10^{8}$ photons at the $15^{th}$ harmonic at
$51.5$~nm are generated per laser shot. The driving intensity is chosen 
on purpose at such a level that the
$15^{th}$ harmonic is positioned exactly at the cut-off of the HHG
process. As a result the next harmonic (the $17^{th}$, causing
background counts due to direct ionization of helium atoms in the spectroscopy 
experiment downstream) is about $10$ times weaker.

\subsection{Spectroscopy chamber}

\begin{figure}[!t]
\includegraphics[width=0.9\columnwidth]{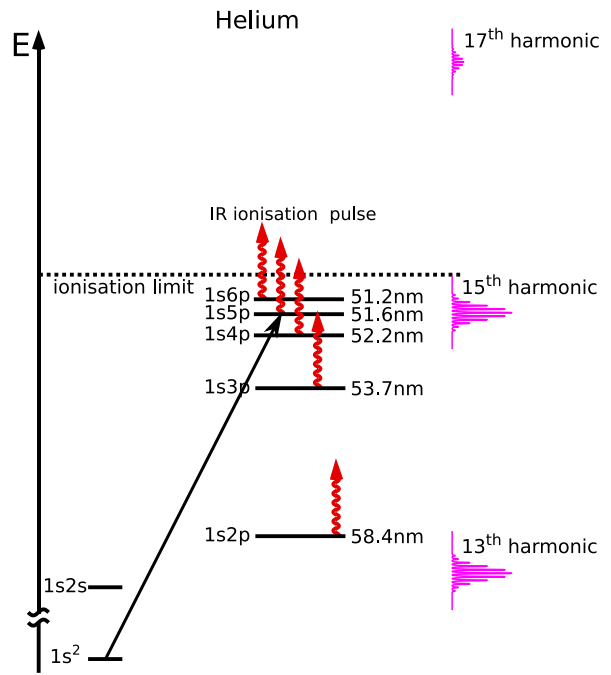}
\caption{The XUV comb drives transitions from the $1s^{2}$~$^{1}\mathrm{S}_{0}$ 
to the $1snp$~$^{1}\mathrm{P}_{1}$ states with the
$15^{th}$ harmonic, and excites the continuum with the
$17^{th}$ and higher harmonics. The excited state population of the $1s4p$ state
and above is selectively ionized using $1064$~nm radiation. Compared
to this schematic, the harmonics are in reality narrower, and contain
many more modes (about $50000$).
\label{spectroscopy_scheme}}
\end{figure}

In the interaction chamber the XUV double pulse intersects a low-divergence 
beam of helium atoms at perpendicular angle to avoid a
Doppler shift. The atomic beam is generated using a pulsed valve
(General Valve, backing pressure $3$~bar) producing a supersonic expansion with either pure
helium or a mixture with a noble gas (ratio $1$~He : $5$~NG). By seeding in Kr, 
Ar and Ne the helium velocity can be varied over a factor of $4$ to
investigate Doppler effects. The divergence of the atomic beam is
limited to approximately $3-4$~mrad by two skimmers: one circular skimmer 
of $0.3$~mm diameter and one adjustable slit skimmer of $0.25$~mm width to set
the XUV-He beam angle. This divergence is similar to the divergence of the XUV
beam ($<2$~mrad). Directly after interaction with the XUV-pulse pair
the excited state population is detected by state-selective ionization (see
Fig.~\ref{spectroscopy_scheme}) with a $50$~ps pulse at $1064$~nm. The
resulting helium ions are detected in a time-of-flight mass
spectrometer.

\section{Frequency metrology on $1s^{2}-1s4p$ and $1s^{2}-1s5p$ transitions in helium}\label{measurement}
The generated XUV comb has been used to measure the $1s^2-1s4p$ and $1s^2-1s5p$ 
transition frequencies, from which an improved value for the helium ground 
state binding energy is derived. Because this involves tests of many
systematic effects, a hierarchy of measurements can be identified (see table~\ref{errors}). The
first level is that of a single recording of the helium 
signal as a function of the repetition rate of the frequency comb
laser in the IR. To record the helium signal, the infrared FC-laser repetition
frequency is scanned in steps of less than $20$~mHz, resulting in
changes in the time separation between the two pulses of around $1$~attosecond. 
Each scan requires recording of $20000-30000$ laser
pulses, which takes about $15$~minutes. For each laser shot, we
record the ion signal, together with a series of 
other parameters, which are the frequency comb
repetition frequency $f_{rep}$ and offset frequency $f_{CEO}$, the
individual IR-pulse energies, energy $E_{XUV}$ of both XUV pulses 
and the average amplifier phase shift. The records are then binned into
typically 20 groups over a Ramsey period, based on a scaled coordinate $u$ defined as
\begin{equation}\label{period}
u:=[q(f_{CEO}+\frac{\overline{\Delta\psi}}{2\pi}f_{rep}) - f_{th}]/f_{rep}
\end{equation}
where $q$ is the harmonic order ($q = 15$), $f_{th}$ is the
theoretically predicted transition frequency, and
$\overline{\Delta\psi}$ is the phase shift at the peak 
of the envelope, which is calculated from the phase measurement and
additional corrections (see section~\ref{discussion}). We calculate the
excitation probabilities $p(u)$ per laser shot, and averages of the
other measured parameters within each bin, denoted by bars over 
the respective symbol from here on. The measured transition
frequency $f_{tf} = f_{th} + f_{ex}$, average background $p_{0}$, and
Ramsey-fringe amplitude $A$ are then fitting parameters in the following model:
\begin{equation}\label{model_eq}
p(u) = \bigg{(}p_{0} + A\cos{2\pi[u + f_{ex}
/\overline{f_{rep}}(u)]}\bigg{)}/\overline{E_{XUV}(u)}
\end{equation}
 
This results in a transition frequency $f_{ex}$ for a single scan, relative to 
the theoretical transition frequency $f_{th}$. The
statistical error in this fit is determined via a bootstrap
method~\cite{Press1992}, which requires no model of the noise sources.

The theoretical transition frequencies $f_{th}$, used in the fitting 
procedure as a reference to which the experimental value is determined,
are obtained by combining recent values of the theoretical ground state energy
from the literature~\cite{Yerokhin2010p022507} with those of the
excited states~\cite{Morton2006p83}. Predicted theoretical frequencies
are $5740806963(36)$~MHz for the $1s^{2}$~$^{1}\mathrm{S}_{0} -
1s4p$~$^{1}\mathrm{P}_{1}$ and $5814248632(36)$~MHz for the $1s^{2}$~$^{1}\mathrm{S}_{0}
- 1s5p$~$^{1}\mathrm{P}_{1}$ transitions.

In Fig.~\ref{wiggle060515} a typical recording of the excitation of
helium by scanning the XUV comb over the $1s^{2}$~$^{1}\mathrm{S}_{0} - 1s5p$~$
^{1}\mathrm{P}_{1}$ transition is shown. The contrast of 
the modulation (in this example $40\%$) is smaller than unity due to various effects, 
like Doppler broadening, frequency noise and a finite constant 
background of about $15\%$ due to direct ionization with the $17^{th}$ and higher
harmonics. Fitting of a single recording typically shows a statistical error of 
$1/50$th of a modulation period. Depending on the repetition
rate it amounts to an uncertainty of $2-3$~MHz, which is
unprecedented in the XUV spectral region. 

Several scans are then performed while changing one parameter
in the setup (e.g. helium velocity) to determine the
magnitude of systematic effects. Such a measurement sequence is
referred to as a ``series'' in the following. The two most frequently performed tests
(series) determine the Doppler shift (by changing the helium
velocity using gas mixtures) and ionization-induced frequency shift 
(by varying the density in the krypton jet). Less frequently the IR-pulse 
ratio was also varied to test its influence on the 
measured transition frequency. The analysis of these tests, as well
as additional experiments that have been performed to determine other
systematic errors, are described in more detail in section~\ref{discussion}. Each of
the series typically requires $4-6$ scans. From this data we extract
(inter/extrapolate) the measured transition frequency in absence of the 
investigated effects and a coefficient for correcting 
the shifted values. For example, the extrapolation 
to zero velocity in a Doppler-shift measurement yields a Doppler-free 
frequency plus the slope, i.e. the angle between XUV and atomic beam. 
The angle can be used to correct the Doppler shift of scans at a finite 
known helium velocity. A ’session’ then consist
of a number ($2-15$) of these series containing in total up to $60$
scans. Within each session, groups of series are selected so that each
group contains at least one Doppler measurement and one
ionization-effect measurement. If possible, also a pulse-ratio-variation 
measurement is included, otherwise a default dependence is
used based on previous measurements. 

\begin{figure}
\includegraphics[width=0.9\columnwidth]{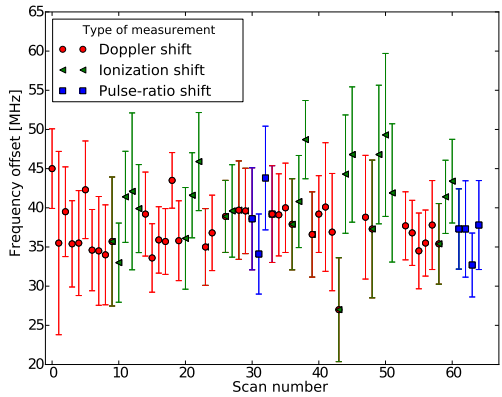}
\caption{A measurement session performed on the $1s^{2}$~$^{1}\mathrm{S}_{0}
- 1s5p$~$^{1}\mathrm{P}_{1}$ transition in helium at 
$f_{rep} = 121$~MHz. The uncorrected frequency offset from theory 
$f_{th}$ is displayed on the y-axis. The different shapes of the points indicate 
measurement series of systematic effects. Doppler-measurement series
are shown as red circles, ionization shift measurement series with green triangles while 
the blue squares denote the measurement of pulse-intensity related effects.
}
\label{121session}
\end{figure}

Figure~\ref{121session} gives an example 
for a data set acquired in a measurement session. The different 
series are displayed therein as red circles for Doppler-shift measurements, 
green triangles for ionization-shift determinations and blue squares for
the measurement of the shift related to the pulse-intensity ratio.
Note, that some measurements are used to derive both the Doppler and 
the ionization-related contribution. These separate determinations are
used to mutually correct each other for the different tested aspects
(so e.g. an ionization-series value is corrected for the Doppler shift
measured just before or after, and vice versa). The error bar on each
series depends on the statistical error of its single recordings, and
the fit of the systematic effect. After grouping, a sequence of
transitions frequencies results with an appropriate error
bar. Additional corrections (such as Stark shift, recoil shift) are
then applied, and the theoretical value for the energy of the excited state is taken into
account. The weighted average over these values then leads to a
session-level value for the ground state binding energy
and an error estimate for it.

Each session is performed for a combination of one upper state in helium
(either $4p$ or $5p$), with one repetition rate of the frequency
comb laser. Most recordings were made of the $1s^{2}$~$^{1}\mathrm{S}_{0} -
1s5p$~$^{1}\mathrm{P}_{1}$ transition. As an additional check also one series was
measured on the $1s^{2}$~$^{1}\mathrm{S}_{0} - 1s4p$~$^{1}\mathrm{P}_{1}$ transition. 
A weighted average over all sessions then leads to a new
value for the measured transition energies. 

\section{Discussion}\label{discussion}

The experiments discussed in this article yield a
two-fold result. The phase coherence of the XUV-pulse pair
allows to measure the frequency of an electronic transition in helium. 
In addition, the contrast of the Ramsey fringes, obtained for different 
repetition frequencies and helium velocities is used to investigate
the coherence of the generated XUV radiation. 
Accordingly the following discussion is divided in two major
parts: an evaluation of the measured transition frequencies and a contrast analysis 
of the observed interference signals.

\subsection{Determination of the ground-state energy of helium with an XUV-frequency comb}

For an absolute measurement
many systematic effects have to be considered in addition to the already 
mentioned effects of Doppler shifts, phase shifts in the NOPCPA and in the HHG. 
These include recoil shifts, refractive index
changes (Kerr effect) in the focusing lens for HHG and the entrance
window to the vacuum setup, and the influence of the plasma generated
in the HHG medium. Also AC and DC Stark effects and Zeeman shifts are
considered. A new ionization energy for helium is derived from these measurements
by taking the excited states ionization energy of the $4p$ and $5p$
into account. The binding energies of these states are 
known to an accuracy better than $20$~kHz from calculations~\cite{Yerokhin2010p022507,Morton2006p83}. 
From the systematic studues an improved experimental 
value is then obtained; an error budget for this value is listed in Table~\ref{errors}

To remove the ambiguity due to the periodic
comb spectrum, the energy of the ground state of helium is determined by measuring the
frequency of $1s^{2}$~$^{1}\mathrm{S}_{0} - 1s5p$~$^{1}\mathrm{P}_{1}$ transition at
different repetition rates $f_{rep}$ of the FCL between $100$
and $185$~MHz. Additionally, the frequency of the $1s^{2}$~$^{1}\mathrm{S}_{0} -
1s4p$~$^{1}\mathrm{P}_{1}$ transition is measured at 
$f_{rep}=148.5$~MHz as a crosscheck. The correct
“mode number” is found by plotting the possible energies of the
helium ground state against $f_{rep}$ as shown in figure~\ref{vernier}, 
(similarly as implemented in~\cite{Witte2005p400}).

Systematic phase and frequency shifts, which occur at different
stages in the experiment (like phase shifts in the NOPCPA or HHG), are
measured, investigated and corrected at different stages of data
analysis. In this section the magnitude and treatment of individual
systematic contributions to the measured frequency is discussed.

Many error sources lead to a frequency uncertainty that is a
function of the delay between the pulses in a Ramsey 
pulse-pair. This can be a direct influence
(by phase shifts), but also indirect (e.g. Doppler shift) because the
fit accuracy depends on the contrast of the Ramsey interference pattern, which in turn
depends e.g. on the repetition rate, helium velocity and decay rate
vs. repetition time. We take these variations into account, providing
not only an error, but also an error range estimate for several
contributions (see Table~\ref{errors}). 
 

Some of the considered error sources relate to differential phase shifts in the IR or XUV
light field (e.g. a IR wavefront error or ionization shift), while
other errors shift the measured frequency (e.g Doppler shift). For
this reason the corrections and errors are given either in terms of a
phase shift at the $15^{th}$ harmonic, or as a frequency shift. Phase
shifts convert into different frequency shifts depending on the
repetition rate of the frequency comb laser, while the frequency
shifts are independent of $f_{rep}$.

In the following sections, we will discuss the effect of differential phase shifts in the amplifier,
in the focusing lens and the entrance window into the HHG-chamber,
and in the krypton jet in which the XUV is generated. Furthermore
Doppler effects, AC and DC Stark shifts, Zeeman shift,
recoil shift and the contribution to the signal from the adjacent
$6p$-level will be evaluated. The discussion is subdivided according to 
the different stages of error analysis following Table~\ref{errors}.

\subsubsection{Single recording level errors}

Unequal conditions experienced by the two laser pulses can shift
their relative phase $\Delta\phi_{CE}$, and therefore the 
offset frequency $f_{CEO}$ of the corresponding frequency comb.  Because every
imprecision in the carrier-envelope phase shift $\Delta\phi_{CE}$
between the IR pulses is multiplied by the harmonic order of the 
harmonic upconversion process, special care must be taken to determine
the resulting phase shifts. The harmonic order of $15$ employed in this
experiment sets a high demand for the detection of differential phase
shifts between the IR pulses. The setup used to
measure the relative spatial-phase profile of the IR-pulse pair was
described in section~\ref{phase}. Typical measured phase shifts are on the
order of $100$~mrad (both positive and negative, depending on the
alignment of the pump laser and NOPCPA system). Therefore in the
$15^{th}$ harmonic phase shifts on the order of
$1.5$~rad in the XUV at $51$~nm are considered. The
phase-shift variation across the spatial profile of the beam is
typically $20-30$~mrad in the IR, which dominates the uncertainty in the 
propagation of the phase shift error to the XUV.

In a previous experiment~\cite{Witte2005p400} the effective influence
of the measured differential spectral phase shift $\Delta\psi(f)$ was
calculated by taking a simple equal weight average of all spectral
components in the pulse. However, this is an oversimplified procedure. 
In order to propagate IR pulse distortions to the XUV we model 
the HHG process using a slowly-varying-envelope approximation, 
neglecting depletion of the HHG medium (consistent with our 
operating conditions). We can then write the (complex) 
generated electric field at the $q^{th}$ harmonic 
\begin{equation}\label{harmonic_field}
E_H(t)=f_H(A(t))e^{iq(\omega_0 t+\phi(t))}+c.c.,
\end{equation}
where $A(t)$ is the slowly varying (real) envelope, $\phi(t)$ 
the (chirp) phase of the pulse and $f_H$ the (complex) single 
frequency response of the HHG medium. We take $\omega_0$ to be 
the $q^{th}$ subharmonic of the transition frequency $\omega_{tr}$. 

The phase shift between the driving laser field and generated XUV
radiation depends on the IR intensity. On a single-atom level and in the
strong-field limit it can be described
analytically~\cite{Lewenstein1995p4747}. This model contains several
contributions to the harmonic yield. The most common of these are
referred to as the ’long’ and ’short’ quantum
trajectories which exhibit a different intensity-dependent linear phase.
By selecting only the central part of the HHG
emission we ensure that only those terms that show the smallest phase
coefficient (the ’short’ trajectory) contribute. The yield of the on-axis
emission from the 'short' trajectory is optimized by placing the HHG medium slightly behind 
the focus of the IR fundamental beam. 
If we now consider the peak of the pulse, assume that only 
a single quantum trajectory contributes to the HHG field (as should 
be the case by design of the experiment) so that there 
is no net interference and therefore
$f_H$ has no oscillatory behaviour, and neglect phase matching 
(which should be fair in a sufficiently small region around the 
peak amplitude), we can model the response function by:
\begin{equation}\label{transfer_function} 
f_H(A) = \alpha A^{n}e^{i\beta A^2} 
\end{equation}
where $\alpha$ and $\beta$ are 
amplitude and phase coefficients and $n$ is an exponent characterizing 
the HHG conversion efficiency. The harmonic field of the 
second pulse with a (possibly) slightly distorted envelope 
$A^*(t) = A(t)+\delta A(t),\quad \phi^*(t)=\phi(t)+\delta\phi(t)$ is then given by
\begin{widetext}
\begin{equation}\label{dist_field}
E^*_H(t)\approx\left(1+\delta A\frac{f_H'(A)}{f_H(A)}+iq\delta\phi\right)f_H(A(t))e^{iq(\omega_0 t+\phi(t))}+c.c.
\end{equation}
so that the Fourier transform of the distorted pulse at the transition 
frequency (neglecting the rapidly oscillating counter-rotating component) reads
\begin{equation}\label{harmonic_spectrum}
\widehat{E^*_H}(\omega_{tr}) \approx \int \dd t \left[1+n\frac{\delta 
A(t)}{A(t)}+i(2\beta A(t) \delta A(t)+q\delta\phi(t))\right] f_H(A(t))e^{iq\phi(t)}.
\end{equation}
\end{widetext}
By adjusting the intensity difference between 
the two pulses we attempt to make the $\delta A$ 
terms vanish. Then the phase difference at the transition frequency 
between the pulses is given by the expression
\begin{equation}\label{long_eqn}
\overline{\Delta \psi} = \arg\left(\frac{\widehat{E_H^*}(\omega_{tr})}{\widehat{E_H}(\omega_{tr})}\right)\approx 
\frac{\int \dd t q\delta\phi(t) f_H(A(t))e^{iq\phi(t)}}{\int \dd t f_H(A(t))e^{iq\phi(t)}}.
\end{equation}
For Fourier-limited pulses ($\phi(t)=0$) and real $f_H$ (no 
intensity-induced phase effect in the HHG process) this would 
just be a weighted average of $\delta\phi$ with $|f_H|$ as a weight function.
However, since the IR pulse is near bandwidth limited and tuned close to 
the transition frequency, $\phi(t)$ can only vary 
a small fraction of unity during the pulse. Likewise, 
because we use a harmonic at the cutoff (the $15^{th}$), 
the XUV yield has a single maximum of limited duration 
around the maximum of the IR pulse. During this time we 
can assume the nonlinear phase $exp(i \beta A^{2})$ to 
be constant to first order (in t). Therefore the phase 
correction required at the transition frequency in the 
XUV is approximately equal to:
\begin{equation}
\overline{\Delta \psi} \approx \frac{\int \dd t q\delta\phi(t) |f_H(A(t))|}{\int \dd t |f_H(A(t))|}.
\end{equation}
Due to the localization of the XUV emission, the weighting function $f_{H}(A(t)))$ could be 
replaced with any function that peaks, where the pulse does.
This means that the effective phase shift in 
the XUV at the transition frequency can be approximated using
phase information in the time domain as the XUV radiation is only
generated when the IR pulse is close to its maximum intensity. We estimate $A(t)$ by 
assuming Fourier-limited IR pulses with the spectrum
found by spectral interferometry. This should 
be a good approximation of the actual pulse shape
in the experiment, as it leads to the highest XUV yield, and the pulse
compressor was set to achieve this maximum XUV yield during the
measurements. $\delta \phi(t)$ is calculated using the same assumption 
by just calculating the inverse Fourier transform of the two 
spectral-interferometry images. In a separate experiment we determined that the
amount of XUV radiation at the $15^{th}$ harmonic depends on the IR
intensity to the $9^{th}$ power, which has actually been used as the 
peaking weighting function for $\delta \phi(t)$. The temporally-averaged phase shift using the
reconstructed XUV intensity and temporal phase shift results in a
value of $\overline{\Delta\psi}$ for each laser shot.  This is then
used in the helium signal fitting procedure.

We analyzed our data using equal weight spectral averaging over the measured spectral phase
as well. The resulting transition frequency coincides with our more
accurate model to within $1$~MHz. However, for the simple spectral
average of the measured phase shift, the spread in the values for the
series with different repetition rates of the comb laser is
significantly larger, which confirms that weighted temporal averaging 
models the phase shift better.

To minimize the intensity-related frequency shifts from HHG in the
spectroscopy, we take care that the energy in the driving pulses is on
average equal within $5\%$. Together with the filtered beam
profile and equal temporal profile, this ensures that the intensity in
the focus is equal on the few percent level. To compensate for possible intensity
differences we determined the observed transition frequency as a
function of IR-pulse energy difference and interpolate linearly to
zero difference. First the energies of the pulses are measured with a 
photo-diode. The integral of the photo-diode signal is recorded 
with an oscilloscope for each pulse and divided by the mean value. 
A value for the peak intensity of a pulse is obtained
from the measured energy, using again the spectral data available
in the interferograms recorded for the phase measurement.
We typically find a frequency shift that, linearly extrapolated, 
corresponds to an IR phase shift smaller 
than $0.2$~rad for $100\%$ relative intensity variation, which is less than 
$3$~rad at the $15^{th}$ harmonic. This interpolation also includes (and therefore
minimizes) nonlinear phase shifts in the beam splitter, focusing lens and vacuum
window, which are not included in the phase measurement using the
Mach-Zehnder interferometer. Our relative intensity 
determination is accurate to within about $2\%$
for small pulse energy differences, so that the error due to this
determination can be estimated to be $60$~mrad, 
corresponding to a frequency error of $1.5$~MHz at $150$~MHz
repetition frequency in the XUV. 

The ``Amplifier phase'' error in Table~\ref{errors} combines the uncertainty of the phase measurement and
wavefront deviation in the fundamental IR beam. The latter is calculated based on an average over the
wavefront scan over the donut mode of the IR pulses. However, in principle there can be a
systematic error for this effect at the level of a single helium signal recording, especially if the
intensity profile is not perfectly homogeneous. In that case the average of the phase measured
over the beam is not representative for the phase shift at the $15^{th}$ harmonic. To minimize
this effect, we employ tight spatial filtering just before the phase-measurement setup to reduce
wave-front errors and to smoothen the intensity profile to the lowest order Gaussian. During a
session the pinhole we use for spatial filtering typically starts to wear out after a few hours of operation, leading to slightly
asymmetric intensity distributions over time that vary randomly for each measurement session.
The pinhole was also regularly replaced and realigned, as was the NOPCPA, so that the intensity and
phase variations average down significantly over the near $200$ scans that were analyzed. Therefore
the error due to the phase deviations and measurement accuracy is listed as a statistical error,
conservatively based on the rms amplitude of the phase deviations as measured for each helium
signal scan.

\subsubsection{Measurement series level errors}

Some of the systematic effects are related to the condition of the setup 
during the measurement. In particular transient and alignment-related effects are 
investigated alternately during a measurement session, forming series of measurements,
which can be used to derive and mutually correct systematic shifts. 

Transient phase shifts occur when the first pulse changes the
propagation conditions for the second one. In this case the condition
of equal pulses is not sufficient to avoid an additional phase shift.
Transient shifts that occur within the Mach-Zehnder interferometer are readily detected and
corrected for. This includes shifts that are caused in the optics
after the NOPCPA.

\begin{figure}
\includegraphics[width=0.9\columnwidth]{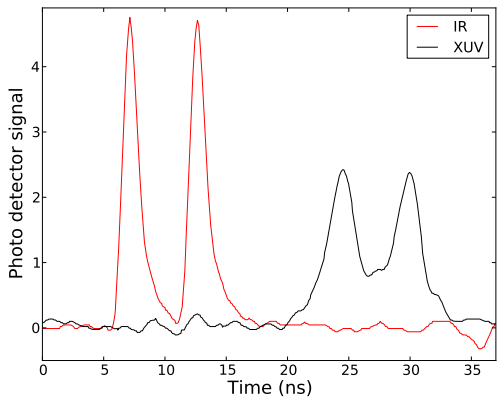}
\caption{(color online) A typical example of the measured IR-pulse photo-detector signals 
(red) together with the corresponding XUV-yield at the $15^{th}$ harmonic (black).}
\label{opaxuv}
\end{figure}

The largest transient effect is due to ionization of the HHG
medium. What matters is only the ionization between the two points in
time where the $15^{th}$ harmonic is generated. Ionization leads to a
lowering of the refractive index especially for the IR fundamental
pulse, to the effect that the phase velocity of the second pulse becomes faster than that of the
first pulse. Ionization is intimately linked to the HHG efficiency and therefore
cannot be avoided completely. As we vary the IR intensity (by
adjusting an iris size in the IR beam path) we find that at too high
intensity, the second XUV pulse can be strongly suppressed due to
ionization of the medium and resulting phase-mismatching effects
induced by the first pulse. In order to keep
the latter to a minimum, we operate at an intensity at which a reasonable
XUV yield is obtained, while simultaneously keeping the XUV 
pulse-energy imbalance to within less than $20\%$ 
(see Fig.~\ref{opaxuv}). In order to determine the remaining
effect due to ionization, we change the pressure in the HHG medium (by
adjusting the valve driver), thereby varying the plasma density, and
record the measured transition frequency effectively for different ion densities 
in the HHG medium. 

The simultaneously measured XUV energy is used as a relative 
gauge for the density of ions in the HHG interaction zone, as 
no direct measurement of the level of ionization of the HHG medium was available for the helium measurements.
If the phase matching conditions are not severely altered due to ionization, 
a quadratic dependence of the XUV output on krypton density 
is expected, while the plasma density, and therefore the 
phase shift, should be proportional to the density of krypton atoms. 
We assume that the induced plasma density 
is proportional to the krypton density, and that no recombination takes place on a timescale of 10 ns.

The relation between XUV yield and ion density was verified 
in a separate experiment by determining the amount of ionization in the HHG process
using a collector grid just below the krypton jet. The voltage applied to 
this grid was on the order of a few volts to avoid 
secondary emission. For the conditions in the experiment the
relation between XUV energy and recorded number of ions is equal to
a power law with an exponent of $2.1(6)$, in good agreement with the
expected $2.0$.

\begin{figure}
\includegraphics[width=0.9\columnwidth]{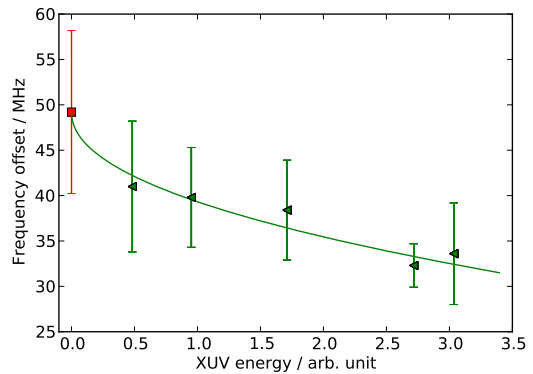}
\caption{A typical measurement of the frequency offset $f_{ex}$ 
(the experimental transition frequency relative to the theoretical 
reference value) as a function of the XUV energy. The XUV energy 
was varied by changing the gas density in the HHG jet.
The green curve is the fitted dependence, while the red square 
indicates $f_{ex}$ for zero gas density in HHG.}
\label{ionifit}
\end{figure}

To obtain the 'ionization free' frequency we extrapolate the frequency 
offset \emph{vs.} XUV energy (when varying Kr density) to zero XUV energy 
using this experimentally determined exponent (see Fig.~\ref{ionifit}).
From the extrapolations we typically find shifts between $0$ and
$2$~rad at the $15^{th}$ harmonic for standard conditions, with
uncertainties in the range between $0.5$ and $2.0$~rad. The average
correction is on the order of $0.1$ XUV cycles. Figure~\ref{ionifit} 
shows the fitting of a typical ionization measurement series, 
from which a value is derived for $f_{ex}$ at zero gas density 
in the HHG jet and therefore at zero XUV intensity.

We also have considered krypton atoms in the HHG medium possibly left
in an excited state after the first pulse. These excited state atoms
would have a different nonlinear susceptibility and therefore produce
a different nonlinear phase shift. Such a shift would not be detected
by reducing the gas density. However, the HHG cutoff energy for such excited
state atoms would lie far below the cutoff energy of the ground state atoms,
as the first excited state is about $10$~eV above the ground
state. Because the cutoff is already set to the $15^{th}$ harmonic for
ground state krypton, this radiation can not be produced by the
excited Kr atoms.

\begin{figure}
\includegraphics[width=0.9\columnwidth]{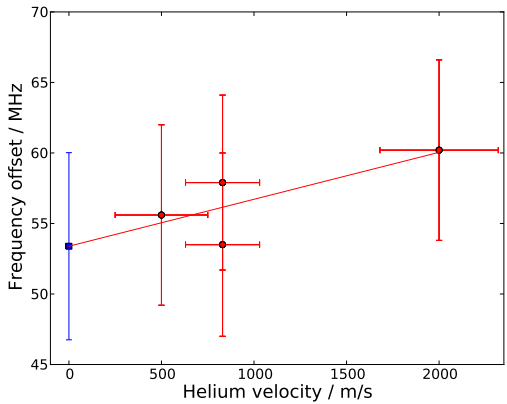}
\caption{A typical measurement of the frequency offset $f_{ex}$ with respect 
to the velocity of the helium atoms in the beam. The red line is the fitted dependence, the blue
square is the extrapolated Doppler-shift free frequency for zero helium velocity.}
\label{dopplerfit}
\end{figure}

The Doppler effect is monitored by varying the helium-beam velocity
using mixtures with noble gases. The velocity of helium in the atomic
beam is measured (in a separate experiment) by monitoring the frequency shift in the helium
signal for different angles between atomic and XUV beam. Helium atoms
in the pulsed atomic beam have a speed of $2000(320)$~m/s for a
pure-helium expansion, helium in neon (pressure ratio 1:5) results in
a speed of $830(200)$~m/s, and helium in argon (pressure ratio 1:5) is
slowed down to $500(250)$~m/s). 
The XUV and atomic beam can be aligned perpendicularly using
the different gas mixtures within approximately $10^{-4}$~rad, resulting in a
typical statistical error of $13$~MHz for a single Doppler shift
determination. 
Figure~\ref{dopplerfit} shows a typical series, measured with pure helium and
the two noble-gas mixtures.
A systematic error in the Doppler shift is introduced 
by the uncertainty in the atomic beam velocity for the different gas mixtures.
Because errors in the He velocity lead to a systematic error in the Doppler 
correction that is proportional to $\cos \phi$ ($\phi$ being the XUV-atomic 
beam angle) and the angle is kept small with changing signs by readjusting 
it, the systematic Doppler shift in the ground state energy is smaller 
than $0.5$~MHz, which is included in the error budget.

\subsubsection{Measurement session level errors}\label{msle}

The biggest correction at the measurement session level comes from 
the recoil shift on the considered transitions and can be
calculated with high precision. It amounts to $18.28$ and $18.75$~MHz for the
$1s$~$^{1}\mathrm{S}_{1} - np$~$^{1}\mathrm{P}_{1} (n \in 4, 5)$ transitions
respectively.

As discussed in section~\ref{HHG}, the IR beam is separated from the XUV beam 
an iris placed after the HHG interaction zone. However, 
some diffracted IR light can still reach the spectroscopy zone. 
Compared to the original IR beam, the diffracted light is found to be at least 
27 times lower in intensity. this light can still produce an AC-Stark shift 
but only during the time of the excitation pulses (and not in between), 
Therefore it shows up in our measurement as a
phase shift, which results in a different frequency shift for
different $f_{rep}$.

The required correction is determined from a measurement with the beam block removed (i.e. with full IR
intensity present in the interaction region). In this case we find a shift of
$17.5(6.0)$~MHz on the $1s^{2}$~$^{1}\mathrm{S}_{0} - 1s5p $~$^{1}\mathrm{P}_{1}$ at
$148$~MHz repetition frequency. With the beam block inserted for the 
regular helium measurements, a shift of less then $0.65$~MHz is expected. 
Converting this to an equivalent phase shift of $14(14)$ mrad in the 
XUV allows to correct the other measurements performed at different
repetition rates. 

We have also estimated the theoretically expected frequency shift at
$f_{rep} =148$~MHz for the same case. The estimated average peak
intensity with the beam stop removed is about $25$~GW/cm$^{2}$, over a pulse duration of
$300$~fs. Between the two pulses the intensity is zero. The AC-Stark
shift can be calculated by taking the time-averaged intensity~\cite{Fendel2007p701} (for
pulses $6.7$~ns apart) of $1$~MW/cm$^{2}$. It leads to a theoretical
AC-Stark shift of $20$~MHz, in good agreement with the experimental
value, measured with the full IR beam passing the interaction zone.

Finally, we need to consider that the pulses for the phase measurement
are split off via a $1$~mm thick beam splitter (with fused silica
as substrate material). After this point the pulses travel through the
focusing lens (few mm BK7 glass) and enter the vacuum setup for HHG
via a Brewster window (another few mm BK7). Any difference in
intensity between the two subsequent pulses that are used for the XUV
comb generation, or any transient effect in these glass pieces, can cause an
additional phase shift which is not taken into account by the generic
phase-measurement procedure during recordings of the helium
transition. To be able to still correct for this phase shift, a
separate experiment was performed removing the IR/XUV separation
iris in the vacuum setup. In this way the IR pulses could travel
through the interaction zone to the monochromator, where they were
coupled out through a window. Two beam splitters inside the
monochromator vacuum chamber reduced the pulse energy to less than $0.1\%$
of the original energy, thus avoiding spurious phase effects in the output
window and other optics. A phase measurement was then performed on
these pulses, while varying the pulse-energy ratio (which can be adjusted via
the pump laser). This measurement resulted in an additional correction
and uncertainty due to the optics placed between the phase-measurement setup 
and HHG (beam splitter, lens and Brewster window) of
$8.5(15)$~mrad in the IR. This contribution (multiplied by 15 
to take the harmonic order into account) is called ``NL-phase shift''
in the error budget in Table~\ref{errors}, where NL stands for the 
nonlinear origin of these differential phase shifts.

From Table~\ref{errors} it is clear that the error contributions differ 
between the sessions due to different sizes of the corresponding data
sets. Also a higher accuracy is achieved for the lower repetition rates in part because phase errors 
there result in smaller frequency errors. The highest uncertainty 
of $25$~MHz at $f_{rep}=185$~MHz is based on a very
short measurement session, consisting of only one set of Doppler and ’ionization’ series.

\subsubsection{Evaluation of the He ground state energy}

In the evaluation of the error in the ground state energy, the combined statistical and systematic 
error based on all sessions ($3.7$~MHz) is combined with several systematic errors represented
as frequencies. 

The biggest contribution to the uncertainty comes from the ionization shift model.
This error contribution listed under ``Ionization shift model'' in Table~\ref{errors} 
refers to the error introduced by the uncertainty in the power law of $2.1(6)$ 
between the XUV intensity and the amount of ionization (and therefore phase shift) in the harmonic 
generation region. As this power law is used to correct for the ionization-induced shifts for
all measurements at the measurement series level, it results 
in a systematic error. This is determined by re-analyzing the helium
ionization potential for a power law of $1.5$ and $2.7$, resulting in a 
variation of $4.9$~MHz. As an additional test of (mostly) ionization-induced 
errors, we also determined the ground state energy using only data points selected
within a limited range of XUV-pulse energy, relative to the average XUV energy for each scan.
Tests were performed for symmetric and anti-symmetric XUV energy distributions with $5\%$ and
$15\%$ deviation from the average XUV energy. A maximum deviation of $5$~MHz (from the value
based on all points) was obtained in case of an asymmetric selection of data points with an XUV
energy higher than $1.15$ times the average energy. This is comparable with the estimated
accuracy of the ionization model as it is based on a typical XUV yield and full distribution of
XUV energies.

During the excitation time interval, the extraction fields for the
time-of-flight spectrometer are switched off. The estimated residual field of less than
$0.5$~V/cm results in a calculated DC-Stark shift on the $5p$,
$m=0$ state of less than $20$~kHz. There can be an additional field due
to ions in the interaction zone left from the first excitation
pulse. However, even for $1000$ ions these fields are too small to
cause a significant shift.

The influence of the Zeeman effect is estimated based on the magnetic 
field measured in the interaction region of $5
\times 10^{-5}$~T. Because the excitation radiation originates from HHG,
which has essentially zero yield for circular polarization, 
we can assume the circular component to be less than $1\%$. 
Since the ground state has no angular momentum, optical 
pumping cannot occur. Therefore only a weak ($1\%$) $\Delta m = \pm 1$ 
component has to be taken into account, leading to a Zeeman shift of less than $7$~kHz.

It is vital for Ramsey excitation, that only one transition is excited
at a time. Excitation of other states would contribute also with a cosine of almost the same period, but a
different phase (which changes quickly with the repetition rate of the
FCL). A contribution from the $1s^{2} - 1s2p$
transition, which is in the vicinity of the $13^{th}$-harmonic 
can be excluded, as it can not be ionized with a single
$1064$~nm photon. But for the excitation of the $5p$ also the $6p$ level could possibly be excited.

\begin{figure}
\includegraphics[width=0.9\columnwidth]{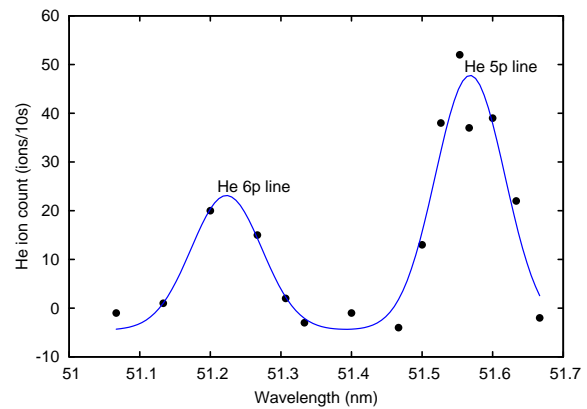}
\caption{Excitation spectrum (single pulse) from the ground state to 
the $1s5p$~$^{1}\mathrm{P}_{1}$ and $1s6p$~$^{1}\mathrm{P}_{1}$ levels 
in helium, obtained by course scanning of the central wavelength
of the fundamental and therefore its $15^{th}$ harmonic over the 
transitions. 
\label{levelscan}}
\end{figure}

To investigate this, the bandwidth of the XUV pulses was determined by coarse tuning of the IR
central wavelength by moving the slit in the Fourier plane of the 
pulse stretcher, located before the parametric amplifier. 
As a result, the XUV comb spectrum is tuned as well, and scanned over the transitions.
From the signal we subtracted a constant background signal from direct
ionization due the $17^{th}$ and higher harmonics. This background has been found by blocking the
ionization beam. A typical scan (corrected for the background) is depicted in Fig.~\ref{levelscan}. 
The solid line is a Gaussian fit of both the $5p$ and the $6p$ resonance, yielding a
1/e half width of the XUV spectrum of $0.07$~nm at the $15^{th}$ harmonic.

From this width and the Gaussian profile we conservatively estimate a contribution to the $5p$
signal from the $6p$ resonance of less than $1\%$. This can cause a shift of the Ramsey pattern of at
most $10$~mrad (in the XUV). However, in practice the influence of this shift on 
the final transition frequency averages down for different repetition
frequencies, so that the maximum error due to excitation of the $6p$ is less than $30$~kHz for all
measurements combined.

\begin{figure}
\includegraphics[width=0.9\columnwidth]{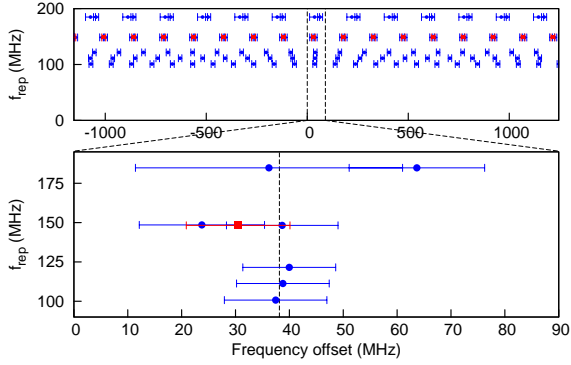}
\caption{Measured $^{4}$He ground state binding energy ($\pm nf_{rep}$ ) relative to the theoretical value of
$5945204174$~MHz~\cite{Yerokhin2010p022507}, plotted against the
repetition frequency $f_{rep}$ of the comb laser. A clear coincidence
is seen at $+37(6)$~MHz relative to theory. The values indicated with
a blue solid circle symbol are based on the $1s^{2}$~$^{1}\mathrm{S}_{0} -
1s5p$~$^{1}\mathrm{P}_{1}$ transition. As a cross check, also a value based on
the recorded $1s^{2}$~$^{1}\mathrm{S}_{0} - 1s4p$~$^{1}\mathrm{P}_{1}$ transition is
included (red square symbol at $f_{rep} = 148$~MHz), resulting in a
consistent ground state energy.
\label{vernier}}
\end{figure}

After incorporating all the systematic corrections, a clear coincidence 
can be seen between the results for different repetition rates. 
A new ground state energy for $^{4}$He of $5945204212(6)$~MHz is found by taking a weighted average over
all measured frequencies at the coincidence location. 
The two most recent theoretical predictions of $5945204174(36)$~MHz 
from~\cite{Yerokhin2010p022507} and $5945204175(36)$~MHz 
from~\cite{Drake2008p45} are in agreement with this value within the
combined uncertainty of theory and experiment. (The uncertainty of $36$~MHz in both theoretical
values is based on estimated but yet uncalculated higher-order QED contributions.) 
Remarkably, very good agreement is found with the prediction of Korobov~\cite{Korobov2001p193003}, 
who calculates $5945204223(42)$~MHz, employing non-relativistic quantum 
electrodynamics theory, in which the problematic divergences of QED are canceled at the operator 
level. However, the uncertainties in the present calculations 
are too large to decide, which of the theoretical approaches is better suited to 
calculate the energy structure of few electron atoms.

Compared with the best previous determinations using single nanosecond laser pulses~\cite{Eikema1997p1866,Bergeson1998p3475}, 
our new value is almost an order of magnitude more accurate. Good agreement is found
with the value of $5945204215(45)$~MHz from~\cite{Eikema1997p1866} 
(based on the measured transition energy in that paper, but 
corrected for a $14.6$~MHz recoil shift that was previously not
taken into account, and using the most recent $2p$ state ionization energy from~\cite{Yerokhin2010p022507}). 
However, there is a difference of nearly $3\sigma$ with the value of $5945204356(50)$~MHz from~\cite{Bergeson1998p3475}.

\begin{table*}
\caption{The error budget for the measurement of
the ground-state binding energy of helium. The corrections and uncertainties are given 
either in radians or in MHz at the $15^{th}$ harmonic, 
depending on the type of effect (see text). The NL-phase shift denotes 
the differential phase shift (multiplied by $15$ to account for its 
effect at the $15^{th}$ harmonic) induced in the optics after the phase 
measurement but before the HHG jet (see~\ref{msle}).\label{errors}}
\begin{tabular}{|l|c|c|c|}\hline
\bf Effect	&\bf Correction	& \bf Systematic uncertainty & \bf Statistical uncertainty	\\ \hline \hline
\multicolumn{4}{|c|}{Single recording level}\\ \hline
Statistical fit error	& ---	& --- & typ.~50--150~mrad  \\ \hline
Amplifier phase	& $0-2.25$~rad	& ---  & $75-375$~mrad \\ \hline
\multicolumn{2}{|l|}{Combined error single recording}	& --- & typ. $160-360$~mrad \\ \hline \hline
\multicolumn{4}{|c|}{Series level}\\ \hline
Doppler shift  & $0 - 10$~MHz	& see text & $8-18$~MHz	\\ \hline
Ionization shift  & $0 - 2$~rad	& see 'ionization shift model' & $0.5 - 2.0$~rad 	\\ \hline \hline
\multicolumn{4}{|c|}{Session level}\\ \hline
AC-Stark shift  & $14$~mrad	& $14$~mrad - see text & --- \\ \hline
NL - phase shift	& $128$~mrad	& $225$~mrad & --- \\ \hline
Recoil shift 4p & 18.28 MHz & $\ll 1$~MHz & --- \\ \hline \hline
Recoil shift 5p & 18.75 MHz & $\ll 1$~MHz & --- \\ \hline 
\multicolumn{2}{|l|}{Combined error single session}	& \multicolumn{2}{|c|}{$9 - 25$~MHz} \\ \hline \hline
\multicolumn{4}{|c|}{Ground state evaluation} \\ \hline
Weighted mean of sessions	& ---	& \multicolumn{2}{|c|}{$3.7$~MHz} \\ \hline
Ionization shift model	& ---	& $4.9$~MHz & --- \\ \hline
Doppler shift	& ---	& $500$~kHz & ---	\\ \hline
DC-Stark shift	& ---	& $< 1$~kHz & --- \\ \hline
Adjacent 6p-level 	& --- & $< 30$~kHz & --- \\ \hline
Zeeman shift	& ---	& $<7$~kHz & --- \\ \hline
\multicolumn{2}{|l|}{Total error in ground state energy}	& \multicolumn{2}{|c|}{$6$~MHz} \\ \hline
\end{tabular}
\end{table*}

\subsection{Signal contrast and phase stability of HHG}

In the previous section phase information of the obtained cosine-shaped signals was used to 
determine the ground state ionization energy of helium. In 
the following we use the observed contrast of the same helium signals 
to investigate the phase stability of the generated XUV frequency comb. The signal visibility 
(ion signal modulation amplitude divided by the average signal) depends on a combination of 
effects. 

\begin{figure}[!ht]
\topbaseline{\includegraphics[width=0.9\columnwidth]{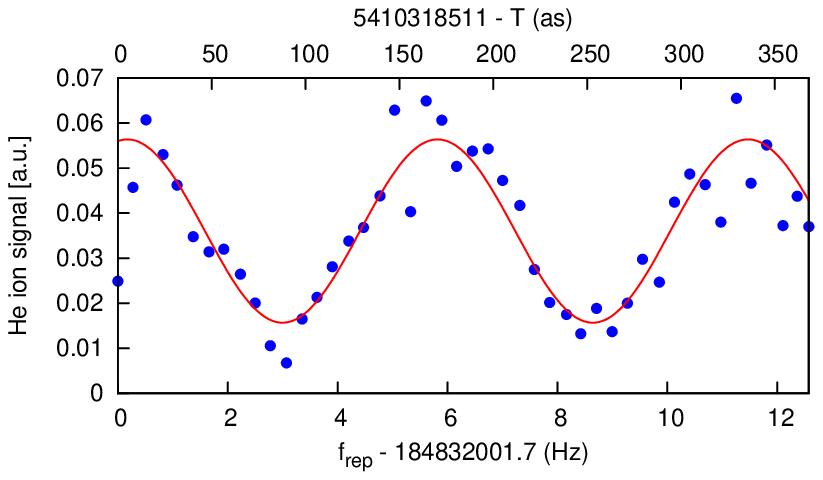}}\llap{\hbox to 0.9\columnwidth{(a)}}
\topbaseline{\includegraphics[width=0.9\columnwidth]{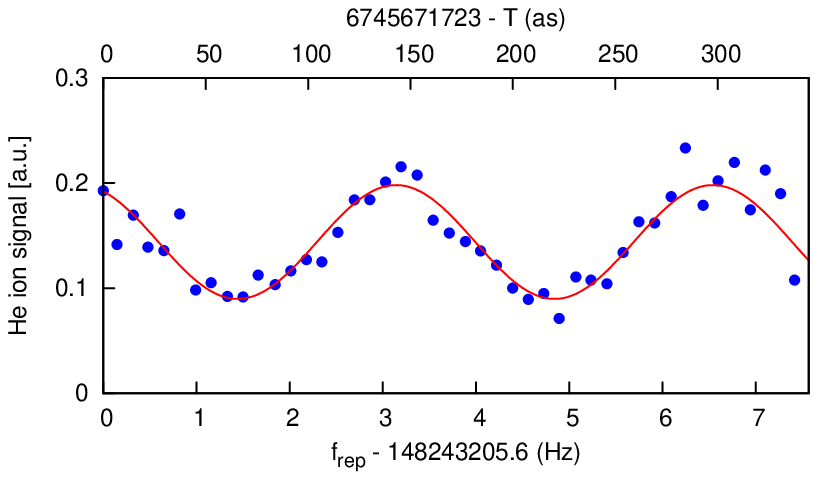}}\llap{\hbox to 0.9\columnwidth{(b)}}
\topbaseline{\includegraphics[width=0.9\columnwidth]{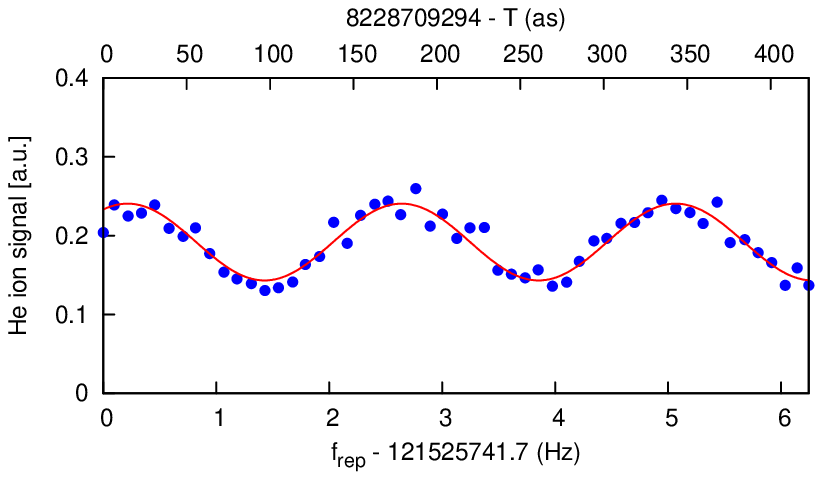}}\llap{\hbox to 0.9\columnwidth{(c)}}
\topbaseline{\includegraphics[width=0.9\columnwidth]{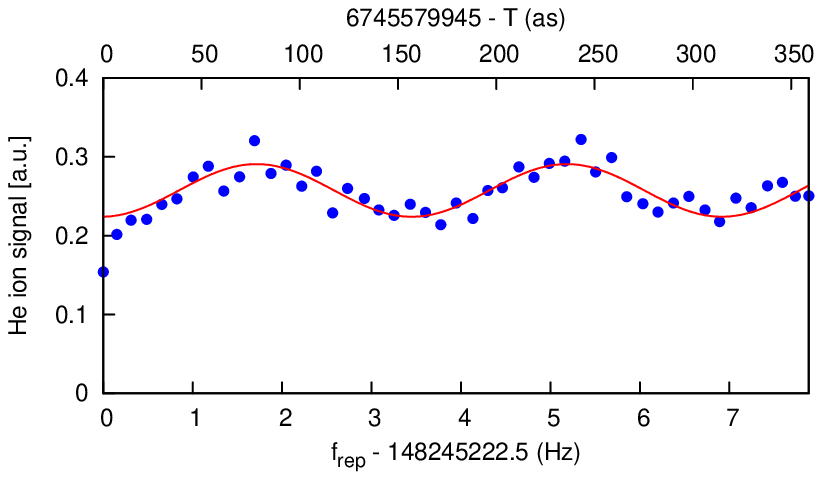}}\llap{\hbox to 0.9\columnwidth{(d)}}
\caption{Ramsey signal of helium, acquired with different experimental conditions. (a) $f_{rep} = 185$~MHz,  $1s^{2}$~$^{1}\mathrm{S}_{0} -
1s5p$~$^{1}\mathrm{P}_{1}$ transition, helium seeded in argon, contrast: $56\%$, (b) $f_{rep} = 148$~MHz,  $1s^{2}$~$^{1}\mathrm{S}_{0} -
1s5p$~$^{1}\mathrm{P}_{1}$ transition, helium seeded in neon, contrast: $38\%$, (c) $f_{rep} = 121$~MHz,  $1s^{2}$~$^{1}\mathrm{S}_{0} -
1s5p$~$^{1}\mathrm{P}_{1}$ transition, helium seeded in neon, contrast: $25\%$, (d) $f_{rep} = 148$~MHz,  $1s^{2}$~$^{1}\mathrm{S}_{0} -
1s4p$~$^{1}\mathrm{P}_{1}$ transition, pure helium beam, contrast: $13\%$.\label{wiggles}}
\end{figure}

The biggest contributions come from the line width of the observed transition, phase stability of the 
$15^{th}$ harmonic, Doppler broadening, and the width of the IR-FCL modes. In one extreme
case ($f_{rep} = 185$~MHz, helium seeded in argon) we find a fringe contrast of $55\%$, while for
$f_{rep} = 100$~MHz and pure helium the contrast is below $5\%$. In Figure~\ref{wiggles}
the signal contrast is shown for several experimental conditions.

From an analysis of the observed visibility of the interference 
pattern as a function of the helium-beam velocity and comb repetition 
frequency, an estimate for the XUV comb phase jitter and of the
effective width of the angular distribution of the atomic beam can be derived. 
We model the Ramsey pattern assuming Gaussian phase noise, while the Doppler 
effect is taken into account with various distributions, including Gaussian and 
rectangular. Both Doppler and phase-noise distribution widths are fitted in the procedure 
to the observed visibilities, and scans are included that were used in the frequency determination, 
plus additional scans that were performed on the $6p$ and $7p$ transitions. In all cases we
assume a common background count rate of $15\%$. The resulting Doppler width and phase jitter
depends on the function used for the Doppler distribution. For a Gaussian profile,
we find a FWHM angular beam divergence of $2.4$~mrad and $0.42$ cycles XUV jitter. A rectangular 
distribution leads to a divergence angle of $4.3$~mrad and $0.35$ cycles XUV phase jitter. In
both cases the model seems to have deficiencies. While in the former case the visibility seems
to be systematically underestimated for slow beams, in the latter case this underestimation is
less pronounced and the fit becomes worse for the fast He beams where the angular distribution
is important. We therefore estimate an intermediate value of $0.38(6)$ cycles for the XUV phase
jitter.

If we consider the contributions to the XUV comb jitter then the biggest 
contribution arises from the bandwidth of the modes of the FC in the IR. 
The bandwidth of the FC was measured 
in a separate experiment by beating the FC modes with 
a frequency doubled narrow bandwidth Er-fiber laser (NP Photonics, 
with a short-term line width of $5$~kHz) to yield a value of 
$1.6$~MHz FWHM, corresponding to a jitter of $1/6$ of an XUV cycle.

A second contribution comes from the phase uncertainty of $47$~mrad 
FWHM in the amplified pulses after correction for the measured phase shift (evaluated
near the peak intensity of the pulse), leading to a jitter of $1/9$ cycle in the XUV. For HHG we
take into account that the phase of the generated harmonics is proportional to the intensity of
the fundamental beam~\cite{Lewenstein1994p2117}, with 
an estimated factor of $\approx10^{-13}$~rad~cm$^{2}$/~W, using the results
presented in~\cite{Corsi2006p023901}. Combined with an 
NOPCPA pulse-ratio distribution FWHM of $2.5\%$, this gives a
small expected jitter of $0.02$ XUV cycles. Fluctuations of $6\%$ in the IR pulse intensity 
lead to variations in the level of ionization in the HHG jet of more than $35\%$ (as ionization
scales with at least the $7^{th}$ power of the IR intensity for the 
conditions in the experiment). If we adopt a typical average ionization
correction of $0.1$ XUV cycles, then this induces a phase jitter of $0.04$ XUV cycles.
Krypton density fluctuations ($<20\%$) induce less than half this 
value. 

Statistically independent combination of these
noise sources leads to a total expected phase jitter in the XUV of $0.21$ cycles. This is lower, but
comparable to the jitter extracted from the contrast measurements. The difference might be due
to uncertainties in the exact experimental conditions, in particular the Doppler broadening and
the bandwidth of the frequency comb laser modes during the experiments. 

The (estimated) contributions to the jitter
suggest that the contrast should hardly be influenced by intensity-induced phase
noise from the HHG process.
This was investigated experimentally by taking only a 
subset of the data points to determine the contrast, 
using a selection criterion based on the measured XUV 
intensity. With this procedure we find that the visibility 
of the interference does not change (within a few
percent) if we restrict the XUV intensity variation to symmetric bands of $5\%$, and $40\%$ around
the mean XUV energy. We therefore conclude that intensity dependent effects do not play a significant role
as a source of phase noise in the XUV comb in the present experiment.


\section{Conclusions and outlook}

\begin{figure}
\topbaseline{\includegraphics[width=0.9\columnwidth]{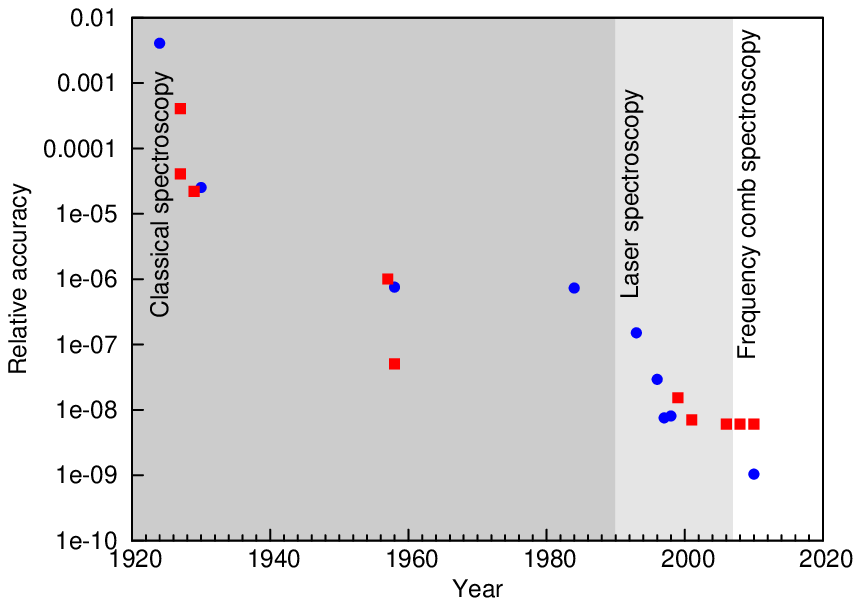}}\llap{\hbox to 0.95\columnwidth{(a)}}
\topbaseline{\includegraphics[width=0.9\columnwidth]{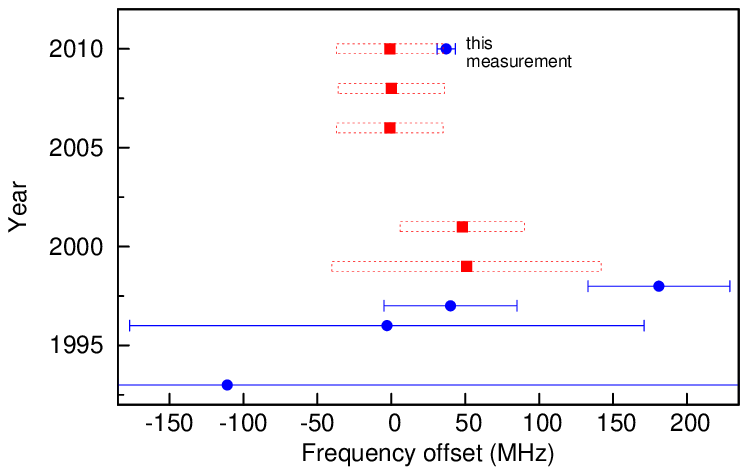}}\llap{\hbox to 0.95\columnwidth{(b)}}
\caption{(color online) (a) Advance in theoretical and spectroscopic accuracy of the
ground state energy of helium over the last 100 years.
The shaded areas emphasize new spectroscopic tools that allowed 
to overcome limits of previous technologies. Red squares 
represent the precision of calculations. The precision obtained 
in experiments is indicated with blue circles.
(b) Comparison of helium ground state ionization energy values 
obtained experimentally (blue circles) and theoretically (red squares) in the last two decades.
The values are plotted with respect to the most recent theoretical 
result~\cite{Bergeson1998p3475,Eikema1997p1866,
Herzberg1958p309,Hylleraas1929p347,Kabir1957p1256,Hopfield1930p133,
Kellner1927p91,Slater1927p423,Yerokhin2010p022507,Baig1984pL383,
Drake1999p83,Drake2008p45,Eikema1993p1690,Eikema1996p1216,
Korobov2001p193003,Lyman1924p1,Pachucki2006p022512,Sucher1958p1010}.
\label{historical}}
\end{figure}

The metrology presented in this article represents a significant advance 
in the history of precision extreme ultraviolet spectroscopy. 
As is illustrated in the historical overview of the spectroscopy 
of the helium atom, depicted in 
Fig.~\ref{historical}, the theoretical and experimental 
accuracy of the value of the binding energy of helium improved by many 
orders of magnitude over the last hundred years. Experimentally 
the biggest progress has always been initiated by new methods, 
which in turn also lead to advances in the theoretical understanding.
The pioneering measurement by Herzberg~\cite{Herzberg1958p309} 
in 1958 based on a helium lamp and grating spectrometer was 
not significantly improved upon until the first laser excitation in 1993~\cite{Eikema1993p1690}. 

In this paper we have described the first absolute frequency 
measurement in the XUV spectral region, based on parametric 
amplification and harmonic upconversion of two pulses 
from an IR frequency comb laser. 
Direct frequency comb excitation in the XUV of 
helium from the ground state is demonstrated, leading to an almost 10-fold 
improved ground state ionization energy, and in good agreement with theory.

For the employed method it is vitally important to control 
and detect phase shifts in the amplification and harmonic 
upconversion process. An important aspect of direct excitation with an upconverted frequency comb is the possibility
to detect systematic errors due to phase shifts between the pulses from the
NOPCPA, HHG and ionization, but also from the AC-Stark effect. Uncorrected errors of this kind show
up as a frequency shift that is proportional to $f_{rep}$, and 
can therefore be detected easily. In the current experiment no 
dependence on the repetition rate is observed in the measurements (see Fig.~\ref{vernier}), 
indicating that the systematic effects have been taken into account properly.
The accuracy of the presented method scales with $f_{rep}$, which 
also means that it can be improved by orders of magnitude by 
choosing frequency comb pulses further apart. If the pulse 
delay would be extended to $> 150$~ns, then the ionization 
shift introduced in the HHG process would in fact vanish 
altogether, as the time between the pulses would be sufficiently long 
to replace the gas in the focal region with a new sample before the second pulse arrives. 
This would reduce the error in the spectroscopy significantly 
because the biggest source of uncertainty is currently due to 
ionization effects in the HHG process. To fully exploit the enhanced
accuracy with larger pulse delays, the sample has to be 
trapped and cooled to reduce Doppler and time-of-flight broadening
which would otherwise wash out the ever denser signal modulation. 
In view of this, measurements on He$^{+}$ are particularly 
interesting. This ion has a hydrogen-like electronic configuration
which can be excited from the ground state using a two-photon transition at $60$~nm. Such an
experiment has the potential to perform QED tests beyond what has been possible so far in
atomic hydrogen \cite{Fischer2004p230802,Herrmann2009p052505}.

In the present work a frequency comb laser in the XUV 
range is demonstrated, for the first time at wavelengths as short as $51$~nm.
The phase jitter of the XUV frequency comb modes is found to be $0.38(6)$ cycles.
From the results we conclude that the observed comb jitter in the XUV is
not dominated by the HHG upconversion process, but most likely by the stability of the original comb in the
infrared. This form of technical noise can be considerably reduced by locking the comb laser
to a stable optical reference cavity. It seems therefore feasible to extend comb generation 
and spectroscopy to even shorter wavelengths into the soft-X-ray region, 
provided that the carrier-phase noise of the fundamental and upconverted 
pulses is kept low enough. Further possibilities would arise if not 
subsequent pulses from the FC, but any two-pulse sequence could be 
selected for upconversion. This would increase the accuracy 
significantly below 1 MHz as the pulses could have a much increased time 
delay, thus making it possible to resolve multiple transitions 
with discrete Fourier transform spectroscopy techniques.
In view of these possibilities, we envision applications such as QED 
test of He$^{+}$ and highly charged ions, precision spectroscopy of 
simple molecules such as $\mathrm{H}_2$, coherent XUV or X-ray imaging, and 
possibly even the emergence of X-ray nuclear clocks.

\bibliography{refbib}
\end{document}